\documentclass[american,aps,prl,twocolumn]{revtex4-1}
\usepackage[T1]{fontenc}
\usepackage[latin9]{inputenc}
\usepackage{amsmath}
\usepackage{babel}
\begin{document}

\title{Envelope modulations of kinetic plasma waves: A general framework}

\author{Ö. D. Gürcan$^{1,2}$, P. Donnel$^{3}$}
\email{ozgur.gurcan@lpp.polytechnique.fr}

\affiliation{$^{1}$CNRS, Laboratoire de Physique des Plasmas, Ecole Polytechnique,
Palaiseau}

\affiliation{$^{2}$ Sorbonne Universités, UPMC Univ Paris 06, Paris}

\affiliation{$^{3}$IRFM, CEA Cadarache, St. Paul Lez Durance}
\begin{abstract}
The general framework of envelope modulations for plasma turbulence
in the presence of strong magnetic fields is discussed. It is shown
that the problem can be formulated using a functional which is equivalent
to the plasma dielectric function for monochromatic waves. Therefore
expanding around its linear wave solutions, one can describe the weakly
nonlinear evolution of waves. The resulting ``quasi-linear'' dielectric
tensor, can describe the evolution of small scale fluctuations advected
and modulated by background flows and corrugations, as well as the
evolution of the large scale flows driven by small scales. As a concrete
example of the framework, gyrokinetic evolution of zonal flows and
drift waves is considered and the resulting nonlinear Schrödinger
equation is derived.

\end{abstract}
\maketitle
Plasma waves, in the form of Zakharov system describing the nonlinear
collapse of Langmuir waves\cite{zakharov:72}, were one of the prime
examples of solitary wave dynamics along with lasers in optical fibers\cite{hasegawa:73},
described using amplitude equations such as the nonlinear Schrödinger
equation (NLS)\cite{sulem-Nls}. The general framework of envelope
modulations from these examples, were rapidly understood to be applicable
to the description of a large class of dispersive nonlinear waves\cite{newell-Book1}.

The basic idea that the index of refraction of the medium in which
a wave propagates, gets modified by the amplitude of the wave itself
applies to all kinds of fluctuations including kinetic plasma waves\cite{goldman:84},
Alfven waves\cite{kaup:78} (in the form of derivative NLS) or drift
waves\cite{champeaux:01} in plasmas. In this latter example the wave
modifies the background medium nonlinearly by driving large scale
(can be termed mesoscale depending on the context) sheared flow structures,
called zonal flows\cite{diamond:05} that advect and shear the small
scale fluctuations that generate them\cite{biglari:90}. This coevolution
of waves and flow structures play a key role in the pattern formation,
in particular by imposing the direction of anisotropy implied by the
nonlinear dynamics of the system\cite{gurcan:15}. 

The application of the general framework as described by A. C. Newell\cite{newell-Book1}
is straightforward to any fluid system. As a result, a number of amplitude
equations were derived for fluid descriptions of various waves and
instabilities in plasma turbulence including drift waves\cite{champeaux:01}
described using simple fluid models\cite{hasegawa:77,hasegawa:83}
as well as ion temperature gradient (ITG) or electron temperature
gradient (ETG) driven instabilities\cite{gurcan:04,gurcan:04b}. 

While modulational instability\cite{chen:00} as well as weakly nonlinear
amplitude equations has been one of the key issues for kinetic plasma
waves. The application of this particular framework to kinetic problems
remain relatively scarce. Here we extend the phenomonology of A. C.
Newell to electrostatic kinetic waves, that can be described by the
gyrokinetic Vlasov equation using a formulation based on the dielectric
tensor.

Consider the dispersion relation for a strongly magnetized electrostatic
plasma wave, which can be written using the plasma dielectric function:
\begin{equation}
\varepsilon\left(\omega,\mathbf{k}\right)\Phi_{\mathbf{k}\omega}=0\;\text{.}\label{eq:bir}
\end{equation}
This relation links the wavevector $\mathbf{k}$ to the complex frequency
$\omega$, where $\Phi$ is the electrostatic potential, and it needs
to be solved for $\omega=\omega_{k}$ in order to obtain a wave solution.
The kinetic dispersion relation is, in general, a complicated function
and has an infinite number of solutions as well as branch cuts. However,
most of these roots have large negative imaginary parts and are therefore
strongly damped. These damped modes can be neglected in the linear
phase, apart from special cases of initial conditions. However as
the system gets into the weakly nonlinear phase, the classification
in terms of ``linearly stable'' and ``linearly unstable'' modes
becomes confused and care must be taken to treat them\cite{terry:06}.

More generally, the plasma dielectric tensor depends on various local
``background'' plasma parameters, such as gradients of temperature,
density, or various components of momentum. In a local linear formulation,
these effects come either from advection by a background flow, or
from the background distribution function $f_{0}\left(X,v,t\right)$.

When these dependencies are explicitly included, we can write:
\begin{equation}
\varepsilon\left(\omega-\mathbf{v}_{E0}\cdot\mathbf{k},\mathbf{k},\partial f_{0}\right)\Phi_{\mathbf{k}\omega}=0\;\text{,}\label{eq:iki}
\end{equation}
where $\mathbf{v}_{E0}$ is the background plasma flow and $f_{0}$
is the background distribution function of, say the ions. The slow
spatial dependence of the background distribution function denoted
by $\partial f_{0}$ define background plasma parameters, such as
gradients of density, momentum and temperature and provides the production
term in the kinetic equation. For kinetic instabilities, velocity
derivatives should also be included (hence the notation $\partial f_{0}$).

With no loss of generality, we can assume that the function $\varepsilon$
correctly represents small and large scale dynamics in all directions.
Note that most problems can be written in this form, taking something
like:
\begin{equation}
\varepsilon\left(\omega,k\right)=\begin{cases}
\varepsilon_{S}\left(\omega,k\right) & k>k_{c},\omega>\omega_{c}\\
\varepsilon_{L}\left(\omega,k\right) & k<k_{c},\omega<\omega_{c}
\end{cases}\label{eq:filt}
\end{equation}
by defining cut-off frequencies $\omega_{c}$ and wavenumbers $k_{c}$.
In practice the cut off may be anisotropic or there may be multiple
bands instead of just two.

The full nonlinear dynamics can be written by considering the following
functional acting on $\Phi$, which can be seen as a generalization
of the plasma dielectric function:
\begin{equation}
\hat{\varepsilon}\left(\partial_{t},\nabla,f_{s}\right)\Phi=0\;\text{,}\label{eq:epsgen}
\end{equation}
where $f_{s}$ is the distribution function corresponding to the species
$s$. Here, the nonlinear term is implicit in the dependence of the
functional on the distribution functions $f_{s}$ and does not appear
as an explicit ``source'' term. In contrast, when (\ref{eq:epsgen})
is ``filtered'' to describe a given scale $k$ (or a finite range
of scales), it becomes an equation, describing the evolution of the
filtered $\Phi_{k}$ with an explicit source term involving other
scales.

Using a standard scale seperation argument and substituting into (\ref{eq:epsgen}),
the expansion
\begin{equation}
\Phi=\widetilde{\Phi}+\overline{\Phi}\quad\text{and}\quad f=f_{0}+\widetilde{f}+\overline{f}\;\text{,}\label{eq:exp-1}
\end{equation}
where the notation $\widetilde{\left(\cdot\right)}$ and $\overline{\left(\cdot\right)}$
depict small scale (faster) and meso scale (slower) perturbations
respectively, and $f_{0}$ is the background profile, we can write:
\begin{equation}
\hat{\varepsilon}\left(\partial_{t},\nabla,f\right)\overline{\Phi}+\hat{\varepsilon}\left(\partial_{t},\nabla,f\right)\widetilde{\Phi}=0\;\text{.}\label{eq:scsp}
\end{equation}
Note that filtering of large and small scales would be done seperately
on the operator and on the fields, since the operator can be different
for large and small scales as discussed above (i.e. Eqn. \ref{eq:filt}).
In other words, if we write (\ref{eq:scsp}) with the large scale
operator (relevant for large scales), we would get $\hat{\varepsilon}_{L}\left(\partial_{t},\nabla,f\right)\overline{\Phi}+\hat{\varepsilon}_{L}\left(\partial_{t},\nabla,f\right)\widetilde{\Phi}=0$,
whereas if we write it for small scales we would get $\hat{\varepsilon}_{S}\left(\partial_{t},\nabla,f\right)\overline{\Phi}+\hat{\varepsilon}_{S}\left(\partial_{t},\nabla,f\right)\widetilde{\Phi}=0$.
Considering the large scale dynamics of (\ref{eq:scsp}) and averaging
(over time, or doing a low pass filter explicitly), we obtain:
\begin{equation}
\hat{\varepsilon}_{L}\overline{\Phi}+\left(\frac{\delta\hat{\varepsilon}_{L}}{\delta f}\delta\overline{f}\right)\overline{\Phi}+\left\langle \left(\frac{\delta\hat{\varepsilon}_{L}}{\delta f}\delta\widetilde{f}\right)\widetilde{\Phi}\right\rangle =0\;\text{,}\label{eq:lpf}
\end{equation}
which is the averaged evolution equation for the large scales including
the large scale self-nonlinearity. Here we used the simplified notation
$\hat{\varepsilon}_{L,S}\equiv\hat{\varepsilon}_{L,S}\left(\partial_{t},\nabla,f_{0}\right)$
for the linear part of the dielectric functional and used variation
in such a way that for instance for a Poisson bracket written as an
operator $\hat{P}\left(f\right)\Phi=\left\{ f,\Phi\right\} $ we can
write:
\[
\left(\frac{\delta\hat{P}}{\delta f}\delta g\right)\Phi\equiv\left\{ g,\Phi\right\} \;\text{.}
\]
High pass filtering the small scale version of (\ref{eq:scsp}), we
can write the equation for the fluctuations:
\begin{equation}
\hat{\varepsilon}_{S}\widetilde{\Phi}+\left(\frac{\delta\hat{\varepsilon}_{S}}{\delta f}\delta\overline{f}\right)\widetilde{\Phi}+\left(\frac{\delta\hat{\varepsilon}_{S}}{\delta f}\delta\widetilde{f}\right)\overline{\Phi}=0\;\text{,}\label{eq:hpf}
\end{equation}
where we dropped the small-scale/small-scale nonlinearity in the spirit
of scale seperation. The system of equations given by (\ref{eq:lpf})
and (\ref{eq:hpf}) can be interpreted as a generalized quasi-linear
system formulated using the form of the dielectric functional.

\paragraph{Small Scales-}

If one considers a given small scale $k$ in (\ref{eq:hpf}), the
first term $\hat{\varepsilon}_{S}$ acting on a plane wave solution
of the form 
\begin{equation}
\widetilde{\Phi}=\widetilde{\Phi}_{\boldsymbol{k}}\left(\boldsymbol{X},T,\tau\right)e^{-i\left(\omega_{k}t-\boldsymbol{k}\cdot\boldsymbol{x}\right)}+c.c.\label{eq:plane_wave}
\end{equation}
produces the linear dielectric function (\ref{eq:iki}) multiplying
the fourier amplitude $\widetilde{\Phi}_{\boldsymbol{k}}$ {[}that
is $\varepsilon_{S}\left(\omega_{\boldsymbol{k}},\mathbf{k},\partial f_{0}\right)\equiv\hat{\varepsilon}_{S}\left(-i\omega_{\boldsymbol{k}},i\mathbf{k},f_{0}\right)${]}.
If a slow spatio-temporal variation is considered, this can be described
by expanding the dielectric functional around $\omega_{\boldsymbol{k}}$
(i.e. one of the relevant roots of the dispersion function) as
\begin{align}
\partial_{t} & \rightarrow-i\omega_{\boldsymbol{k}}+\epsilon\partial_{T}+\epsilon^{2}\partial_{\tau}\nonumber \\
\nabla & \rightarrow i\boldsymbol{k}+\epsilon\nabla_{X}\;\text{.}\label{eq:ops}
\end{align}
where $\epsilon$ is a smallness parameter. Note that, even though
it is not shown explicitly here, one has to track how the operator
acts on the complex conjugate in (\ref{eq:plane_wave}) as well. For
the linear propagator, we can see the above expansion also as an expansion
of the dielectric function (\ref{eq:iki}) with $\omega\rightarrow\omega_{\boldsymbol{k}}+i\epsilon\partial_{T}+i\epsilon^{2}\partial_{\tau}$
and $\mathbf{k}\rightarrow\boldsymbol{k}-i\epsilon\nabla$. However
this requires changing our view of the dielectric function, since
the result of the expansion will be an operator acting on $\Phi$.
Since our motivation here is to see where slow modulations of the
amplitude will be balanced by the first non-negligible nonlinear terms,
we include the modifications due to slowly evolving meso-scale flows
and corrugations as well. The operators acting on meso-scale quantities
$\overline{f}$ and $\overline{\Phi}$ will give $\epsilon\partial_{T}$
and $\epsilon\nabla_{X}$ directly. Furthermore, since the amplitudes
can be assumed to be small initially (and growing) until a point where
the modulations of the linear term can be balanced by a nonlinear
term, we can choose particular scalings of $\widetilde{f}$, $\widetilde{\Phi}$,
$\overline{f}$ and $\overline{\Phi}$ as long as their relations
are consistent with the equations that link them (e.g. quasi-neutrality)
in order to derive a weakly nonlinear amplitude equation.

Notice that when $\overline{\Phi}$ and $\overline{f}$ are set to
zero and a single plane wave solution is considered for $\widetilde{\Phi}$
as in (\ref{eq:plane_wave}) one recovers the linear dielectric function
(\ref{eq:iki}). This means that the expansion (\ref{eq:exp-1}) can
also be seen as an expansion of the linear dispersion function (\ref{eq:iki})
by replacing $\partial f_{0}\rightarrow\partial f_{0}+\partial\overline{f}+\partial\widetilde{f}$
and $v_{E0}\rightarrow\hat{b}\times\nabla\overline{\Phi}$. 

Writing (\ref{eq:epsgen}) order by order using (\ref{eq:ops}) and
(\ref{eq:exp-1}), and taking $\widetilde{\Phi}$ as in (\ref{eq:plane_wave})
we get:

\begin{equation}
O\left(1\right):\quad\varepsilon\left(\omega_{\boldsymbol{k}},\boldsymbol{k},\nabla f_{0}\right)\widetilde{\Phi}_{\boldsymbol{k}}=0\;\text{,}\label{eq:o0}
\end{equation}
which is true since $\omega=\omega_{\boldsymbol{k}}$ is a solution
of $\varepsilon\left(\omega,\boldsymbol{k},\nabla f_{0}\right)=0$,
the linear dispersion relation. In the next order we find:

\begin{equation}
O\left(\epsilon\right):\quad\left.i\frac{\partial\varepsilon\left(\omega,\boldsymbol{k},\nabla f_{0}\right)}{\partial\omega}\right|_{\omega=\omega_{k}}\left(\partial_{T}+\frac{\partial\omega_{\boldsymbol{k}}}{\partial\boldsymbol{k}}\cdot\nabla\right)\widetilde{\Phi}_{\boldsymbol{k}}=0\;\text{,}\label{eq:O1-2}
\end{equation}
which basically describes group velocity propagation of the wave packet.
Note that in order to obtain (\ref{eq:O1-2}), we have used $\frac{\partial\varepsilon}{\partial\boldsymbol{k}}=-\frac{\partial\omega_{\boldsymbol{k}}}{\partial\boldsymbol{k}}\frac{\partial\varepsilon}{\partial\omega}$,
which can be obtained by taking the total derivative of (\ref{eq:o0})
with respect to $\boldsymbol{k}$. Expanding up to the second order,
we obtain:
\begin{align}
O\left(\epsilon^{2}\right):\;\bigg\{\frac{\partial\varepsilon}{\partial\omega_{k}}\big(i\partial_{\tau}+\frac{1}{2}\frac{\partial^{2}\omega_{\boldsymbol{k}}}{\partial k_{i}\partial k_{j}}\partial_{ij} & -\boldsymbol{k}\cdot\overline{\boldsymbol{v}}_{E}\big)\nonumber \\
-\frac{1}{2}\frac{\partial^{2}\varepsilon}{\partial\omega_{k}^{2}}\left(\partial_{T}+\frac{\partial\omega_{k}}{\partial\boldsymbol{k}}\cdot\nabla\right)^{2} & +\frac{\partial\varepsilon}{\partial\left[\nabla f_{0}\right]}\cdot\nabla\overline{f}\bigg\}\widetilde{\Phi}_{\boldsymbol{k}}=0\label{eq:O2-1}
\end{align}
 where we used the shorthand notation $\frac{\partial\varepsilon}{\partial\omega_{k}}=\left.\frac{\partial\varepsilon\left(\omega,\boldsymbol{k},\nabla f_{0}\right)}{\partial\omega}\right|_{\omega=\omega_{k}}$
and $\frac{\partial\varepsilon}{\partial\left[\nabla f_{0}\right]}=\left.\frac{\partial\varepsilon\left(\omega,\boldsymbol{k},\nabla f_{0}\right)}{\partial\left(\nabla f_{0}\right)}\right|_{\omega=\omega_{k}}$,
and the second derivative of (\ref{eq:o0}) in order to simplify (\ref{eq:O2-1}).

Thus, the full equation for the evolution of the fluctuations under
the action of large scale flows and corrugations, can be written in
general as:
\begin{align}
\bigg\{ i\left(\partial_{T}+\frac{\partial\omega_{k}}{\partial\boldsymbol{k}}\cdot\nabla\right)+\epsilon\bigg[i\partial_{\tau}+\frac{1}{2}\frac{\partial^{2}\omega_{\boldsymbol{k}}}{\partial k_{i}\partial k_{j}}\partial_{ij}\nonumber \\
-\boldsymbol{k}\cdot\overline{\boldsymbol{v}}_{E}-\frac{a}{2}\left(\partial_{T}+\frac{\partial\omega_{k}}{\partial\boldsymbol{k}}\cdot\nabla\right)^{2}+\boldsymbol{b}\cdot\nabla\overline{f}\bigg]\bigg\}\Phi_{\boldsymbol{k}} & =0\label{eq:smsc1}
\end{align}
where 
\[
a\equiv\frac{\partial^{2}\varepsilon}{\partial\omega_{k}^{2}}\big/\frac{\partial\varepsilon}{\partial\omega_{k}}\quad\text{and}\quad b_{i}\equiv\frac{\partial\varepsilon}{\partial\left[\partial_{i}f_{0}\right]}\big/\frac{\partial\varepsilon}{\partial\omega_{k}}
\]
are functions of $k$.

Note that, one can switch to the frame of reference that is moving
with the group velocity of the fluctuations, in which (\ref{eq:smsc1})
becomes simply
\begin{equation}
\bigg[i\partial_{\tau}+\frac{1}{2}\frac{\partial^{2}\omega_{\boldsymbol{k}}}{\partial k_{i}\partial k_{j}}\partial_{ij}-\boldsymbol{k}\cdot\overline{\boldsymbol{v}}_{E}+\boldsymbol{b}\cdot\nabla\overline{f}\bigg]\Phi_{\boldsymbol{k}}=0\label{eq:smsc2}
\end{equation}

\paragraph{Large Scales-}

In order to close (\ref{eq:smsc1}), we need the evolution equations
for $\overline{f}$ and $\overline{\Phi}$ (i.e. $\overline{\boldsymbol{v}}_{E}=\hat{\boldsymbol{b}}\times\nabla\overline{\Phi}$),
which can be obtained by substituting the expansion of the operators
(\ref{eq:ops}) and the equivalent expansions 
\begin{align}
\partial_{t} & \rightarrow\epsilon\partial_{T}+\epsilon^{2}\partial_{\tau}\nonumber \\
\nabla & \rightarrow\epsilon\nabla_{X}\;\text{.}\label{eq:ops-1}
\end{align}
for the large scales into $\hat{\varepsilon}_{L}$ in (\ref{eq:lpf}).
At this point we can decide to keep or drop the large scale self nonlinearity
represented by the second term in (\ref{eq:lpf}). If we want to obtain
an amplitude equation for the evolution of small scale envelopes due
to its self-consistent interactions with large scales, we have to
``solve'' the large scale dynamics represented by the first term
in (\ref{eq:lpf}) in terms of the small scale stresses represented
by the last term. In order to do that we have to linearize (\ref{eq:lpf})
in terms of the large scale fields. This gives: 
\[
\hat{\varepsilon}_{L}\overline{\Phi}+\left\langle \left(\frac{\delta\hat{\varepsilon}_{L}}{\delta f}\delta\widetilde{f}\right)\widetilde{\Phi}\right\rangle =0\;\text{,}
\]
Assuming that the $\hat{\varepsilon}_{L}$ can be inverted for example
by performing a Fourier transform, we can write
\[
\overline{\boldsymbol{v}}_{E}=\hat{\boldsymbol{b}}\times\nabla\left[\hat{\varepsilon}_{L}^{-1}\left\langle \left(\frac{\delta\hat{\varepsilon}_{L}}{\delta f}\delta\widetilde{f}\right)\widetilde{\Phi}\right\rangle \right]
\]
Looking back at (\ref{eq:smsc1}), we also need $\overline{f}$. Unfortunately
the general formulation in terms of the dielectric functional does
not allow us to obtain such an equation. We can either go back to
the original kinetic equation that led to the definition of $\hat{\varepsilon}_{L}$,
or equivalantly for the single species case we can assume a linear
relation between $\overline{\Phi}$ and $\overline{f}$ consistent
with dropping large scale nonlinear terms. However one interesting
observation allows us to sidestep this troublesome issue altogether.
When the ion dynamics is considered, the usual large scale limit would
be nicely aligned with the magnetic field lines (i.e. large scale
$\nabla_{\parallel}\rightarrow0$). As the electrons can not respond
to perturbations with $\nabla_{\parallel}=0$, the quasi-neutrality
condition gives
\begin{equation}
\left[1-\Gamma_{0}\left(b\right)\right]\overline{\Phi}=\int\overline{f}J_{0}v_{\perp}dv_{\perp}dv_{\parallel}\label{eq:qnzf}
\end{equation}
where $\Gamma_{0}$ is the modified Bessel function. When (\ref{eq:qnzf})
is expanded with $b=\nabla_{\perp}^{2}$ by taking $\nabla_{\perp}\rightarrow\epsilon\nabla_{\perp}$
one finds that $O\left(\overline{f}\right)=\epsilon^{2}O\left(\overline{\Phi}\right)$
and therefore the $\overline{f}$ term appearing in (\ref{eq:smsc1})
and (\ref{eq:smsc2}) can be dropped. The resulting amplitude equation
in the group velocity frame, can be written as:
\begin{align}
\bigg\{ & i\partial_{\tau}+\frac{1}{2}\frac{\partial^{2}\omega_{\boldsymbol{k}}}{\partial k_{i}\partial k_{j}}\partial_{ij}\nonumber \\
 & -\boldsymbol{k}\cdot\hat{\boldsymbol{b}}\times\nabla\left[\hat{\varepsilon}_{L}^{-1}\left\langle \left(\frac{\delta\hat{\varepsilon}_{L}}{\delta f}\delta\widetilde{f}\right)\widetilde{\Phi}\right\rangle \right]\bigg\}\Phi_{\boldsymbol{k}}=0\;\text{.}\label{eq:pre_nls}
\end{align}
Note that, this equation has the structure of a nonlinear Schrödinger
equation, even though the exact structure of the nonlinear term depends
on the nonlinear structure of the dielectric functional.

\paragraph*{NLS for Gyrokinetic Drift Waves: Zonal Flows-}

Considering local gyrokinetic description of toroidal ITG\cite{coppi:67,kim:94,kuroda:98}
and zonal modulations one can obtain the weakly nonlinear evolution
of drift waves and zonal flows. The operator form of the dielectric
tensor, for non-zonal modes can be written as:
\begin{equation}
\hat{\varepsilon}_{S}\left(\partial_{t},\nabla;f\right)=\left(1+\frac{1}{\tau}-\Gamma_{0}\right)\left[\cdot\right]-\int\mathcal{L}_{f}^{-1}\left\{ J_{0}\left[\cdot\right],f\right\} J_{0}d^{3}v\label{eq:eps_s_itg}
\end{equation}
where $\tau=T_{e}/T_{i}$ is the ratio of electron to ion temperatures
and the gyrokinetic equation is:
\[
\mathcal{L}_{f}\delta f=\left\{ J\Phi,f\right\} 
\]
with $\delta f=f-f_{0}$ and $\mathcal{L}_{f}=\left(\partial_{t}+v_{\parallel}\nabla_{\parallel}+\hat{\omega}_{D}\left(v\right)\right)$.
Here $\hat{\omega}_{D}\left(v\right)$ is the curvature drift and
$J$ is the guiding center averaging operator, which turns into the
Bessel function $J_{0}\left(k_{\perp}v_{\perp}\right)$ in $k$-space.
Similarly
\begin{equation}
\hat{\varepsilon}_{L}\left(\partial_{t},\nabla;f\right)=\left(1-\Gamma_{0}\right)\left[\cdot\right]-\int\left(\mathcal{\overline{L}}_{f}\right)^{-1}\left\{ J_{0}\left[\cdot\right],f\right\} \overline{J}_{0}d^{3}v\label{eq:eps_l_itg}
\end{equation}
with $\mathcal{\overline{L}}_{f}=\left(\partial_{t}+\nu\right)$ and
$\overline{J}_{0}\approx1$ is the gyroaverage operator on a zonal
perturbation.

Linearizing, one obtains the dispersion relation, which can be solved
using curvature modified plasma dispersion functions\cite{gurcan:14}.
When $\eta_{i}\equiv d\ln T_{i}/d\ln n_{0}<2/3$, the ITG is damped,
but the drift wave branch is a stable wave. Thus, if one takes a plane
drift wave solution in this range, one can apply the methodology discussed
above.

In order to compute the nonlinear term, we expand the Poisson bracket
as
\begin{align*}
\left\{ J_{0}\widetilde{\Phi},\delta\widetilde{f}\right\}  & =\left\langle \hat{\boldsymbol{b}}\times\nabla J_{0}\widetilde{\Phi}\cdot\nabla\delta\widetilde{f}\right\rangle \\
 & =J_{0k}^{2}\left(\hat{\boldsymbol{b}}\times\boldsymbol{k}\cdot\nabla\right)\left(\frac{d\chi_{k}}{d\boldsymbol{k}}\cdot\nabla\right)\left|\widetilde{\Phi}_{k}\right|^{2}
\end{align*}
after rather tedious algebra, where
\[
\delta\widetilde{f}_{k}=\mathcal{L}_{fk}^{-1}\left\{ J_{0k}\widetilde{\Phi}_{k},f_{0}\right\} =-\chi_{k}J_{0k}\widetilde{\Phi}_{k}
\]
is used in order to define the coefficient $\chi_{k}=\chi_{k}\left(v\right)$
as a partial linear response of $\delta\widetilde{f}_{k}$ to $\widetilde{\Phi}_{k}$.
The nonlinear term that appears in (\ref{eq:pre_nls}) becomes:
\begin{align*}
N.L= & -\bigg[\hat{\varepsilon}_{L}^{-1}\left(\mathcal{\overline{L}}_{f}\right)^{-1}\\
 & \times\int J_{0k}^{2}\left(\hat{\boldsymbol{b}}\times\boldsymbol{k}\cdot\nabla\right)^{2}\left(\frac{d\chi_{k}}{d\boldsymbol{k}}\cdot\nabla\right)d^{3}v\left|\widetilde{\Phi}_{k}\right|^{2}\bigg]\widetilde{\Phi}_{k}\;\text{.}
\end{align*}

The operator $\hat{\varepsilon}_{L}\mathcal{\overline{L}}_{f}$ ,
can have two different limiting forms for zonal modulations. For the
case $\nu<\epsilon$ it is the group velocity term that dominates
and we get $\hat{\varepsilon}_{L}\mathcal{\overline{L}}_{f}\approx-v_{gx}\partial_{XX}\partial_{X}$,
which gives a standard NLS equation:
\begin{equation}
\left(i\partial_{\tau}+\frac{1}{2}\frac{\partial^{2}\omega_{\boldsymbol{k}}}{\partial k_{x}\partial k_{x}}\partial_{XX}\right)\Phi_{\boldsymbol{k}}+\beta_{k}\left|\widetilde{\Phi}_{k}\right|^{2}\Phi_{\boldsymbol{k}}=0\;\text{,}\label{eq:nls}
\end{equation}
where $\beta_{k}=\frac{1}{v_{gx}}\int J_{0k}^{2}k_{y}^{2}\frac{d\chi_{k}}{dk_{x}}d^{3}v$.
The existence of soliton structures such as those implied from (\ref{eq:nls})
in toroidal ITG is interesting. One may speculate that weakly coherent
modes observed in tokamaks (e.g. Ref \cite{manz:15}) may be linked
to such structures. Also the obvious connection to wave kinetics,
allows the study of the wave number spectrum and the condansate\cite{connaughton:05}.

A general framework for the description of kinetic plasma waves in
terms of a plasma dielectric function is presented, which can be applied
to various problems including the zonal flow/drift wave system resulting
in a NLS equation. It is probably worth noting that one of the future
perspectives of this work is the extension of the method to electromagnetic
waves.
\begin{acknowledgments}
Part of this work was done in the framework of the projects programme
of \emph{Festival de Theorie} in \emph{Aix en Provence} 2017.
\end{acknowledgments}

\end{document}